\documentclass[twocolumn,preprintnumbers,amsmath,amssymb]{revtex4}
\usepackage{graphicx}% Include figure files
\usepackage{dcolumn}% Align table columns on decimal point
\usepackage{bm}% bold math

\begin{document}

%\preprint{APS/123-QED}
%
\newcommand{\bea}{\begin{eqnarray}}
\newcommand{\eea}{\end{eqnarray}}
\newcommand{\be}{\begin{equation}}
\newcommand{\ee}{\end{equation}}
%
%%%%%%%%%%%%%%%%%%  BOLDFACE GREEK LETTERS   %%%%%%%%%%%%%%%%
\newcommand{\xbf}[1]{\mbox{\boldmath $ #1 $}}
%%%%%%%%%%%%%%%%%%%%%%%%%%%%%%%%%%%%%%%%%%%%%%%%%%%%%%%%%%%%%

\title{Excited nucleon electromagnetic form factors from broken spin-flavor
 symmetry~\footnote{12$^{th}$ Int'l. Workshop on the 
Physics of Excited Nucleons, NSTAR 2009, 
Beijing, April 19-22, 2009, Chin. Phys. C {\bf 33}, 1257 (2009).}}

\author{A. J. Buchmann}
\affiliation{Institute for Theoretical Physics \\
University of T\"ubingen \\
D-72076 T\"ubingen, Germany}
\email{alfons.buchmann@uni-tuebingen.de}

\begin{abstract}
%%%%%%%%%%%%%%%%%%%%%%%%%%%%%%%%%%%%%%%%%%%%%%%%%%%%%%%%%%%%%%%%
A group theoretical derivation of a relation between 
the $N \to \Delta$ charge quadrupole transition and neutron charge
form factors is presented.
\end{abstract}
%%%%%%%%%%%%%%%%%%%%%%%%%%%%%%%%%%%%%%%%%%%%%%%%%%%%%%%%%%%%%%%%%

\maketitle

\section{Introduction}
A milestone in the development of strong interaction theory 
was the proposition~\cite{Gel64} 
that strong interactions not only conserve isospin and strangeness but are 
also approximately invariant under the higher SU(3) flavor symmetry.
The latter combines baryon isospin multiplets with different isospin $I$ and 
strangeness $S$ to larger degenerate
multiplets of particles with the same spin ${\bf J}$ and parity $P$,
e.g. to a baryon octet with spin 1/2  and a baryon decuplet with spin 3/2.
Although broken, flavor symmetry leads to a number of remarkable
predictions such as the Gell-Mann-Okubo relation for octet baryons
and the equal spacing rule for decuplet baryons, both of which are well 
satisfied in nature.

A still higher symmetry is obtained
when SU(3)$_F$ flavor and SU(2)$_J$  spin symmetries are 
embedded into the larger SU(6) spin-flavor group, in which case 
new generators link the previously unconnected flavor 
and spin symmetries~\cite{Gur64,Sak64,Beg64}.  
The assumption of an underlying spin-flavor symmetry of strong interactions 
has far greater predictive 
power than individual flavor and  spin symmetries
represented by the direct product group SU(2)$_J \times$SU(3)$_F$.
For example, within SU(6) it is not only possible
to combine the spin 1/2 flavor octet (2$\times$ 8 states) 
and spin 3/2 flavor decuplet (4$\times$ 10 states) 
into a {\bf 56} dimensional spin-flavor supermultiplet,
but also to connect observables of different spin tensor rank 
such as charge radii (rank 0) 
and quadrupole moments (rank 2) that remain unrelated by 
the direct product group.

Numerous successes, for example 
$\mu_p/\mu_n=-3/2$ for the ratio of proton and neutron magnetic 
moments~\cite{Beg64}, affirm that SU(6) is a useful
symmetry in baryon physics.
We now understand that the underlying field theory of strong interactions,
quantum chromodynamics (QCD), possesses a spin-flavor symmetry which
is exact in the large $N_c$ limit~\cite{sak84,das95}, 
where $N_c$ denotes the number of colors, 
and that for finite $N_c$ spin-flavor symmetry 
breaking operators can be classified 
according to the powers of $1/N_c$ associated with them.
As a result, one obtains a rigorous energy scale independent 
perturbative expansion scheme for QCD processes~\cite{leb98}. 

Previously, we abstracted from the quark model 
with two-body exchange currents 
a relation between the inelastic $N \to \Delta$
quadrupole and the elastic neutron charge form factors~\cite{Buc04}
\bea
\label{ffrel2}
G_{C2}^{N \to \Delta}(Q^2) &  = &  -\frac{3\,\sqrt{2}}{Q^2} G_C^n(Q^2),
\eea
which in the limit of zero photon momentum transfer reduces to 
a relation between the $N \to \Delta$ transition quadrupole moment
and the neutron charge radius~\cite{Buc97} 
\be
\label{ffrel3}
Q_{N \to \Delta} = \frac{1}{\sqrt{2}}\, r_n^2.
\ee
Here, $ N \to \Delta$ stands for both $p \to \Delta^+$ and
$n \to \Delta^0$ transitions.
Comparison of these relations with experiment shows good agreement
from low to high momentum transfers~\cite{dre07}.

It has been pointed out that the derivation of 
Eq.(\ref{ffrel2}) relies only on the 
spin-flavor structure of the wave functions and operators involved, i.e., 
only on general algebraic properties of the quark model and not 
on specific assumptions, such as values for quark masses, coupling 
constants, etc. Therefore, it appears that it should be 
derivable within the framework of an abstract SU(6) tensor analysis, 
similar to the derivation of the Beg-Lee-Pais relation
$\mu_{p \to \Delta^+} = 2\sqrt{2}\mu_p/3$  
between the $p \to \Delta^+$ transition and proton magnetic moments.   
The purpose of this paper is to present a group theoretical derivation
of Eq.(\ref{ffrel3}). The generalization to finite momentum transfers 
will be presented elsewhere.

\section{Spin-flavor symmetry analysis}

%%%%%%%%%%%%%%%%%%%%%%%%%%%%%%%%%%%%%%
%SU(6) spin-flavor symmetry analysis
%%%%%%%%%%%%%%%%%%%%%%%%%%%%%%%%%%%%%% 
We start from the observation that the $N(939)$ and $\Delta(1232)$
are members of the same {\bf 56} dimensional SU(6) ground state multiplet
of spin-flavor symmetry. If the symmetry were exact,
$N$ and $\Delta$ baryons would have the same mass.
Spin-dependent operators in the Hamiltonian $H$ break SU(6)
symmetry and lift the degeneracy between $N$ and $\Delta$
masses. Moreover, they connect the symmetry breaking
in the flavor octet to the symmetry breaking in the flavor decuplet.

Similarly, in the SU(6) symmetry limit the charge form factors
$G_C^n(Q^2)$ and $G_{C2}^{N \to \Delta}(Q^2)$ are exactly
zero.  In the following we will see that
spin-dependent terms in the charge operator $\rho$ 
break SU(6) symmetry and lead to nonzero neutron and $N \to \Delta$
charge form factors, which are related as in Eq.(\ref{ffrel2}) because
the group algebra connects the breaking of the symmetry in
$G_C^n(Q^2)$ to the symmetry breaking in $G_{C2}^{N \to \Delta}(Q^2)$.

A basic assumption in a group-theoretical analysis is that quantum mechanical
operators and states have definite transformation properties, i.e.,
they transform according to certain irreducible representations (reps) of 
the underlying symmetry group.   A general matrix element 
${\cal M}$ of an operator $\Omega^{R}$ 
evaluated between baryon ground states reads 
\be
{\cal M}= \langle {\bf 56} \vert \, 
\Omega^{R} \, \vert {\bf 56} \rangle, 
\ee
where $R$ is the dimension of the irreducible rep associated with
the considered operator. 

An allowed symmetry breaking 
operator $\Omega^{R}$ acting on the baryon 
ground state multiplet must transform 
according to one of the irreducible reps $R$ 
contained in the direct product~\cite{Sak64}
\be
\label{directproduct}
\bar{{\bf 56}} \times {\bf 56} 
=  {\bf 1} + {\bf 35} + {\bf 405} + {\bf 2695}.
\ee 
Here, the ${\bf 1}$ dimensional rep on the right-hand side 
corresponds to an SU(6) symmetric operator, while the remaining reps
characterize respectively, first, second, and third order SU(6) symmetry
breaking operators.
Operators transforming according to other SU(6) reps
not contained in this product
will lead to vanishing matrix elements when evaluated between 
states belonging to the ${\bf 56}$ dimensional ground state multiplet.

In terms of quark degrees of freedom, one 
can think of these operators as being constructed 
from quark-antiquark bilinears transforming according to 
the adjoint {\bf 35} dimensional rep of SU(6), 
arising from the direct product of two fundamental reps 
$ {\bf 6} \times {\bf {\bar 6}} 
= {\bf 1}+ {\bf 35}$.
Then, on the right-hand side of 
Eq.(\ref{directproduct}), the {\bf 1} is associated 
with a zero-quark operator (constant), and the ${\bf 35}$, ${\bf 405}$, and 
${\bf 2695}$, are respectively connected with one-, two-, 
and three-quark operators~\cite{leb95}. 

First order SU(6) symmetry breaking operators, 
i.e., one-quark operators,
which are constructed from the 35 generators of the SU(6) 
group~\cite{comment1},
do not lift the degeneracy between $N$ and $\Delta$ 
masses and do not generate nonzero neutral baryon charge radii and
nonvanishing baryon quadrupole moments.

In order to split the supermultiplet and to differentiate between 
spin 1/2 flavor octet and spin 3/2 flavor decuplet masses, 
the SU(6) symmetry breaking part of $H$ must be spin-dependent. 
On the other hand,  the Hamiltonian transforms as an overall spin scalar.
Therefore, we need at least second order SU(6) 
symmetry breaking operators, i.e., in the simplest case, scalar products  
of two SU(2)$_J$ generators in order for $H$ to satisfy the two conditions of 
being spin-dependent and a spin tensor of rank 0 at the same time.

Analogously, to explain the nonzeroness of the neutron charge 
and decuplet quadrupole form factors 
it is required that the charge operator $\rho$ 
be spin-dependent. 
At the same time, for Coulomb multipoles of $\rho$ 
we need spin tensor operators of even rank
in order to satisfy the time reversal and parity invariances 
of the electromagnetic interaction.
Thus, again at least second order SU(6) symmetry breaking operators are
required for these observables.

Second order SU(6) symmetry breaking two-quark operators 
can be constructed from direct products of one-quark operators.
A general two-quark spin-flavor operator transforms according 
to one of the irreducible reps found in the direct 
product ${\bf 35}\times {\bf 35}$~\cite{Sak64}
\be
\label{gsu6tbo} 
{\bf 35} \times {\bf 35} =  {\bf 1} + {\bf 35} + {\bf 35} + 
{\bf 189} + {\bf 280} + {\bar {\bf 280}} + {\bf 405}.
\ee
Two-quark operators transforming according to the ${\bf 1}$  
and ${\bf 35}$ dimensional SU(6) reps can be reduced to constants 
and one-quark operators in spin space, so that only the four higher 
dimensional reps 
on the right hand side of Eq.(\ref{gsu6tbo}) remain.
Of these only the ${\bf 405}$ rep appears in the direct baryon ground state 
product $\bar{{\bf 56}} \times {\bf 56}$ according to Eq.(\ref{directproduct}).
Therefore, within the ${\bf 56}$ an allowed two-quark operator must 
necessarily transform according to the ${\bf 405}$ dimensional rep
of SU(6). 
 
To have a better understanding of the type of operators involved 
we perform a multipole expansion of the relevant 
two-quark charge density $\rho_{[2]}$ in spin-flavor space 
up to quadrupole terms
\bea
\label{su6break}
\rho_{[2]} &  = & a \, {\cal S}_{[2]} 
+ b {\cal T}_{[2]} , 
\eea
where the spin scalar ${\cal S}_{[2]}$ and spin tensor ${\cal T}_{[2]}$ 
operators are defined as
\bea
{\cal S}_{[2]}  & =  &-B \sum_{i \ne j} \, e_i \, 
\bm{\sigma}_i \cdot \bm{\sigma}_j,  \nonumber \\
{\cal T}_{[2]}  & =  & -B \sum_{i \ne j} \, e_i \, 
(3 \, \sigma_{i\,z} \sigma_{j\,z}
- \bm{\sigma}_i \cdot \bm{\sigma}_j).
\eea
Here, the constant $B$ parametrizes the color 
and orbital matrix elements,  and $e_i=(1 + 3\,\tau_{3\, i})/6$ 
is the quark charge. Furthermore, 
$\bm{\sigma}_i$ and 
$\bm{\tau}_i$ are the spin and isospin Pauli matrices of the i-th quark. 
To first order flavor breaking, 
Eq.(\ref{su6break}) represents the most general two-quark charge 
operator in spin-flavor space. These tensors have been
used in calculating the {\bf 56} baryon ground state charge radii 
and quadrupole moments~\cite{hen02,hes02,buc02,buc07}. 

We will see shortly that as a consequence 
of the underlying SU(6) spin-flavor symmetry, the
spin scalar and spin tensor terms in Eq.(\ref{su6break}) 
have fixed relative strengths $a/b=-2$.
An evaluation of Eq.(\ref{su6break}) between $N$
and $\Delta$ spin-flavor wave functions leads then 
straightforwardly to the following results
\bea
\label{results}
r_n^2 & = &   4 \, B \, a \nonumber \\
Q_{N \to \Delta} & = & 2 \sqrt{2} \, B \, a, 
\eea
from which Eq.(\ref{ffrel3}) is readily established.
\newpage

Next, without reference to the quark model, 
we show that the spin tensors of rank 0 and 2 in 
Eq.(\ref{su6break}) are different
components of a general SU(6) tensor of dimension ${\bf 405}$
which are linked to each other by the group algebra.
A decomposition of the tensor $\Omega_{\bf 405}$ into subtensors 
with definite transformation properties 
with respect to the flavor and spin subgroups of SU(6) reads
\bea
\label{405decomp}
{\bf 405} 
& = & 
({\bf 1},{\bf 1}) + 
({\bf 8},{\bf 1}) + 
({\bf 27},{\bf 1}) 
\nonumber \\
& + &  
2 \,  ({\bf 8},{\bf 3}) 
+  ({\bf 10},{\bf 3}) +  
 ({\bar {\bf 10}},{\bf 3}) +  
 ({\bf 27},{\bf 3}) \nonumber \\ 
& + &  ({\bf 1},{\bf 5}) + 
 ({\bf 8},{\bf 5}) + 
 ({\bf 27},{\bf 5}),  
\eea
where the first and second entry in the parentheses refers to the dimensions
of the SU(3)$_F$ and SU(2)$_J$ representations respectively~\cite{beg64a}.
Thus, spin-flavor symmetry breaking proceeds along the chain 
$SU(6) \supset SU(3)_F \times SU(2)_J \supset SU(2)_I \times U(1)_Y$,
where in a first step SU(6) symmetry is broken 
into SU(3)$_F\times$SU(2)$_J$, and in a second step SU(3)$_F$ symmetry 
is reduced to SU(2)$_T\times$U(1)$_Y$, i.e., an uncorrelated product
of isospin and hypercharge symmetries.  
 
For Coulomb multipoles, we are restricted 
to spin tensors of even rank, and the second line in Eq.(\ref{405decomp}) 
need not concern us here. 
Furthermore, we confine ourselves to flavor octet tensors 
appropriate for electromagnetic interaction operators.
Eq.(\ref{405decomp}) shows that there is 
a unique spin scalar ${\cal S}$ transforming as   
$({\bf 8},{\bf 1})$ and a unique spin tensor ${\cal T}$ 
transforming as $({\bf 8},{\bf 5})$ and both are united in a common SU(6) 
tensor with dimension ${\bf 405}$. 

We expand the baryon charge density operator $\rho$ 
into Coulomb multipoles~\cite{def66} up to quadrupole terms
\bea
\rho \!\!\!\!  & = & \!\!\! \!  
\sum_J i^J \,{\hat J} \, \, T_0^{CJ} = \rho^{C0} + \rho^{C2},
\eea
with ${\hat J}=\sqrt{2J+1}$ and where we have suppressed the momentum 
dependence of the multipole operators and an overall factor $\sqrt{4\pi}$. 
The Coulomb multipole operators $T_0^{CJ}$
are irreducible tensors of rank $J$ and correspond to 
the second order SU(6) symmetry breaking tensors 
$\Omega^{{\bf405}}_{(\mu\, s)}$. We can now identify 
\bea
\rho^{C0} & \sim & \phantom{-\sqrt{5}} \, \, 
\Omega^{\bf 405}_{({\bf 8}, {\bf 1})}
\nonumber \\
\rho^{C2} & \sim &  -\sqrt{5} \, \,
 \Omega^{\bf 405}_{({\bf 8}, {\bf 5})}.
\eea 
The two-quark operators ${\cal S}_{[2]}$ and ${\cal T}_{[2]}$ 
in Eq.(\ref{su6break}) have just the same transformation properties 
and are recognized here as different components of a common ${\bf 405}$ 
dimensional tensor operator $\Omega^{\bf 405}$. 

According to the generalized Wigner-Eckart theorem, 
the matrix elements of $\Omega^{\bf 405}$ evaluated between 
the ${\bf 56}$ multiplet can be factorized into 
a common reduced matrix element (indicated by a double bar), 
which is the same for the entire 
multiplet, and an SU(6) Clebsch-Gordan (CG) coefficient
\bea 
{\cal M} & = & 
\langle {\bf 56}_{\nu_f} \vert \, 
\Omega^{{\bf 405}}_{\nu} \, \vert {\bf 56}_{\nu_i} \rangle \nonumber \\
&=& 
\langle {\bf 56} \vert \vert \, 
\Omega^{{\bf 405}} \, \vert \vert {\bf 56} \rangle \, \, 
\left (\begin{matrix}
{\bf 56} &  {\bf 405} & {\bf 56} \\
\nu_i & \nu & \nu_f
\end{matrix} \right ). 
\eea
The latter provide relations between the matrix elements of 
different components of the irreducible tensor operator 
$\Omega^{\bf 405}_{\nu}$ and the individual states of the
${\bf 56}$ dimensional baryon ground state supermultiplet, 
which are labelled by $\nu_{i}$ and $\nu_{f}$. 
Because SU(6) is a rank five group, the label $\nu$ comprises 
five quantum numbers to uniquely specify a state, three
for SU(3), e.g. total isospin $T$, isospin projection $T_z$,
and hypercharge $Y$, and two for SU(2), e.g. total angular momentum $J$
and its projection $J_z$.

The SU(6) CG coefficient can be split into a unitary scalar 
factor $f_{(\mu,s)}^{{\bf 405}}$ 
and a product of SU(3)$_F$ and SU(2)$_J$ CG coefficients as
\bea 
\left (\begin{matrix}
{\bf 56} &  {\bf 405} & {\bf 56} \\
\nu_f & \nu & \nu_i
\end{matrix} \right )
& = & f^{{\bf 405}}_{(\mu, \, s)}  
\left (\begin{matrix}
\xbf{\mu}_f &  \xbf{\mu} & \xbf{\mu}_i \\
\rho_f & \rho & \rho_i 
\end{matrix} \right ) \nonumber \\
& & \left(J_i\, J_{i,\, 3} J \, J_{3} \vert J_f \, J_{f,\, 3} \right ),
\eea
where $\mu$ and $s=2J+1$ denote the dimensionalities of the
SU(3) and SU(2) reps. The SU(3)$_F$ CG coefficient label 
$\rho$ comprises the three quantum numbers 
$\rho=(Y, T, T_z)$. Note that the 
SU(6) scalar factor $f^{{\bf 405}}_{(\mu, s)}$, 
depends only on the dimensionalities
of the SU(6), SU(3)$_F$, and SU(2)$_J$ reps involved but not on the SU(3) 
and SU(2) labels $\rho$ and $J_z$.

Now, consider the two SU(6) matrix elements, which are of interest here
\bea 
\label{grouptheory}
r_n^2 &  = & \langle {\bf 56}_{n} \vert \, 
\Omega^{{\bf 405}}_{({\bf 8},\,{\bf 1}) } \, \vert {\bf 56}_{n} 
\rangle  \nonumber \\
& = &  r \ \left ( -\frac{2}{\sqrt{10}} \right ) \, 
\left \lbrack \frac{1}{\sqrt{3}}\,\left (-\sqrt{\frac{1}{20}} \right ) 
-\sqrt{\frac{3}{20}}  \right \rbrack \nonumber \\
&= & r \, \, \frac{2\sqrt{6}}{15}, 
\nonumber \\
Q_{p \to \Delta^+} &  =  & -\sqrt{5} \,\, 
\langle {\bf 56}_{\Delta^+} \vert \, \Omega^{{\bf 405}}_{({\bf 8},\,{\bf 5}) } 
\, \vert {\bf 56}_{p} \rangle  \nonumber \\
&=& (-\sqrt{5}) \, r \left ( \frac{1}{\sqrt{10}} \right )
\left \lbrack \frac{2}{\sqrt{15}} \right \rbrack 
\left (- \frac{2}{\sqrt{10}} \right ) \nonumber \\
&= & r \, \, \frac{2\sqrt{3}}{15}, 
\eea
where $r=\langle {\bf 56} \vert \vert \, 
\Omega^{{\bf 405}} \, \vert \vert {\bf 56} \rangle $
is the SU(6) reduced matrix element. The factor of -2 
between the rank 0 (charge monopole) and rank 2 (charge quadrupole) 
tensors is reflected by the SU(6) scalar factors~\cite{coo65,leb95} 
$f^{{\bf 405}}_{(8,1)}=-2/\sqrt{10}$ and $f^{{\bf 405}}_{(8,5)}=1/\sqrt{10}$. 
The SU(3)$_F$ flavor~\cite{mcn65} and SU(2)$_J$ spin CG coefficients
are explicitly shown. In the case of the neutron charge radius, 
the two terms in the brackets correspond to SU(3) CG with 
sublabels $\rho=(0,0,0)$ and $\rho=(0,1,0)$. As usual, the 
isosinglet part of a flavor octet charge operator is multiplied 
by $1/\sqrt{3}$. From Eq.(\ref{grouptheory}) we obtain Eq.(\ref{ffrel3}).
We could have arrived at this result much faster by noting that the 
spin-isospin Clebsch-Gordan coefficients have already been calculated 
in Eq.(\ref{results}) so that the SU(6) scalar factor would have sufficed
to establish Eq.(\ref{ffrel3}).

\section{Summary}

%%%%%%%%%%%%%%%%%%%%%%%%%%%%%%%%%%%%%%
%Summary
%%%%%%%%%%%%%%%%%%%%%%%%%%%%%%%%%%%%%% 

We have seen that for the present application to the neutron charge radius 
and the $N \to \Delta$ quadrupole moment, where first order SU(6) 
symmetry breaking  does not contribute, the $J=0$ and $J=2$ 
multipole components of the charge density $\rho$ transform as the
second order SU(6) symmetry breaking tensors 
$\Omega^{\bf 405}_{(8,1)}$ and $\Omega^{\bf 405}_{(8,5)}$ respectively.
In contrast to separate SU(2)$_J$ and SU(3)$_F$ symmetries or their direct 
product SU(2)$_J\times $SU(3)$_F$, broken spin-flavor SU(6) symmetry 
provides a definite relation between spin operators 
of different tensor rank that belong to the same SU(6) tensor. 
As a result we obtain the relation
between the neutron charge radius and the $N\to \Delta$ quadrupole moment
of Eq.(\ref{ffrel3}) from a general SU(6) symmetry analysis. 
We hope to address a generalization of this derivation including 
third order SU(6) symmetry breaking in a future communication.

\vspace{-2mm}
\centerline{\rule{80mm}{0.1pt}}


\begin{thebibliography}{90}
\bibitem{comment1} 
For one particle, the 35 generators  of SU(6) are composed of
three pure spin operators
$\xbf{\sigma}_i/2$ with $i=1,2,3$,
eight pure flavor operators $\xbf{\lambda}_k/2$ with $k=1,\cdots,8$,
and 24 combined spin-flavor operators
$(\xbf{\sigma_i}/2) \, (\xbf{\lambda}_k/2)$.
\end{thebibliography}
\end{document}